\documentclass[aps,prl,preprint,tightenlines,superscriptaddress,showpacs,byrevtex]{revtex4}
\usepackage{graphicx} 
\usepackage{dcolumn}  

\graphicspath{{ps}}


\newcommand{\ee}{e^{+}e^{-}}
\newcommand{\leplep}{\ell^{+}\ell^{-}}
\newcommand{\jp}{J/\psi}

\newcommand{\ecp}{\eta_{c}(2S)}
\newcommand{\psip}{\psi^{\prime}}
\newcommand{\chic}{\chi_{c1}}

\newcommand{\pipi}{\pi^{+}\pi^{-}}
\newcommand{\bbar}{B$-$\bar{B}}
\newcommand{\kpi}{K^-\pi^{+}}
\newcommand{\ks}{K_{S}}

\newcommand{\Mbc}{M_{\rm bc}}
\newcommand{\DE}{\Delta E}

\newcommand{\rt}{\rightarrow}

\newcommand{\etal}{\em et al.}

\begin{document}


\title{ Observation of a narrow 
charmonium-like state in exclusive $B^{\pm}\rt K^{\pm} \pipi\jp$ decays}

\affiliation{Budker Institute of Nuclear Physics, Novosibirsk}
\affiliation{Chiba University, Chiba}
\affiliation{University of Cincinnati, Cincinnati, Ohio 45221}
\affiliation{University of Frankfurt, Frankfurt}
\affiliation{Gyeongsang National University, Chinju}
\affiliation{University of Hawaii, Honolulu, Hawaii 96822}
\affiliation{High Energy Accelerator Research Organization (KEK), Tsukuba}
\affiliation{Hiroshima Institute of Technology, Hiroshima}
\affiliation{Institute of High Energy Physics, Chinese Academy of Sciences, Beijing}
\affiliation{Institute of High Energy Physics, Vienna}
\affiliation{Institute for Theoretical and Experimental Physics, Moscow}
\affiliation{J. Stefan Institute, Ljubljana}
\affiliation{Kanagawa University, Yokohama}
\affiliation{Korea University, Seoul}
\affiliation{Kyungpook National University, Taegu}
\affiliation{Institut de Physique des Hautes \'Energies, Universit\'e de Lausanne, Lausanne}
\affiliation{University of Ljubljana, Ljubljana}
\affiliation{University of Maribor, Maribor}
\affiliation{University of Melbourne, Victoria}
\affiliation{Nagoya University, Nagoya}
\affiliation{Nara Women's University, Nara}
\affiliation{National Kaohsiung Normal University, Kaohsiung}
\affiliation{National Lien-Ho Institute of Technology, Miao Li}
\affiliation{Department of Physics, National Taiwan University, Taipei}
\affiliation{H. Niewodniczanski Institute of Nuclear Physics, Krakow}
\affiliation{Nihon Dental College, Niigata}
\affiliation{Niigata University, Niigata}
\affiliation{Osaka City University, Osaka}
\affiliation{Osaka University, Osaka}
\affiliation{Panjab University, Chandigarh}
\affiliation{Peking University, Beijing}
\affiliation{Princeton University, Princeton, New Jersey 08545}
\affiliation{RIKEN BNL Research Center, Upton, New York 11973}
\affiliation{Saga University, Saga}
\affiliation{University of Science and Technology of China, Hefei}
\affiliation{Seoul National University, Seoul}
\affiliation{Sungkyunkwan University, Suwon}
\affiliation{University of Sydney, Sydney NSW}
\affiliation{Tata Institute of Fundamental Research, Bombay}
\affiliation{Toho University, Funabashi}
\affiliation{Tohoku Gakuin University, Tagajo}
\affiliation{Tohoku University, Sendai}
\affiliation{Department of Physics, University of Tokyo, Tokyo}
\affiliation{Tokyo Institute of Technology, Tokyo}
\affiliation{Tokyo Metropolitan University, Tokyo}
\affiliation{Tokyo University of Agriculture and Technology, Tokyo}
\affiliation{Toyama National College of Maritime Technology, Toyama}
\affiliation{University of Tsukuba, Tsukuba}
\affiliation{Utkal University, Bhubaneswer}
\affiliation{Virginia Polytechnic Institute and State University, Blacksburg, Virginia 24061}
\affiliation{Yokkaichi University, Yokkaichi}
\affiliation{Yonsei University, Seoul}
  \author{S.-K.~Choi}\affiliation{Gyeongsang National University, Chinju} 
  \author{S.~L.~Olsen}\affiliation{University of Hawaii, Honolulu, Hawaii 96822} 
  \author{K.~Abe}\affiliation{High Energy Accelerator Research Organization (KEK), Tsukuba} 
  \author{T.~Abe}\affiliation{High Energy Accelerator Research Organization (KEK), Tsukuba} 
  \author{I.~Adachi}\affiliation{High Energy Accelerator Research Organization (KEK), Tsukuba} 
  \author{Byoung~Sup~Ahn}\affiliation{Korea University, Seoul} 
  \author{H.~Aihara}\affiliation{Department of Physics, University of Tokyo, Tokyo} 
  \author{K.~Akai}\affiliation{High Energy Accelerator Research Organization (KEK), Tsukuba} 
  \author{M.~Akatsu}\affiliation{Nagoya University, Nagoya} 
  \author{M.~Akemoto}\affiliation{High Energy Accelerator Research Organization (KEK), Tsukuba} 
% ...
\collaboration{The Belle Collaboration}
\noaffiliation

\begin{abstract}

We report the observation of a narrow charmonium-like state produced
in the exclusive decay process $B^{\pm}\rt K^{\pm}\pipi\jp$.  This state, 
which decays
into $\pipi\jp$, has a mass of $3872.0\pm 0.6 {\rm (stat)} \pm 0.5{\rm 
(syst)}$~MeV, a value that
is very near the $M_{D} + M_{D^*}$ mass threshold.  
The results are based on an analysis of 152M $\bbar$ events
collected at the $\Upsilon(4S)$ resonance in the Belle detector at the 
KEKB collider. The signal has a  statistical significance 
that is in excess of $10\sigma$.

\end{abstract}

\maketitle


A major experimental issue for the
$c\overline{c}$  charmonium particle
system is the existence of as yet unestablished charmonium states that 
are expected to be
below threshold for decays to open charm and, thus, narrow.
These include the $n=1$ singlet~P state, the 
$J^{PC}=1^{+-}~1^1{\rm P}_{c1}$, and
possibly the $n=1$ singlet and triplet spin-2~D states, i.e. 
the $J^{PC}=2^{-+}~1^1{\rm D}_{c2}$ and
$J^{PC}=2^{--}~1^3{\rm D}_{c2}$, all of 
which are narrow if their masses are below
the $D\bar{D^*}$ threshold.  The
observation of these states and the determination of
their masses would provide useful information about
the spin dependence of the charmonium potential.

In addition to charmonium states, some authors have
predicted the existence of $D^{(*)}\bar{D}^{(*)}$
``molecular charmonium'' states~\cite{voloshin} and 
$c\overline{c}g$
``hybrid charmonium''
states~\cite{godfrey}.  If such states
exist with masses below the relevant open charm threshold,
they are expected 
to be narrow and to have large branching fractions to
low-lying $c\overline{c}$ charmonium states.

The large $B$ meson samples produced at $B$-factories provide
excellent opportunities to search for new charmonium states.
The Belle group recently reported the first observation of the
$\ecp$ via its $\ks\kpi$ decay channel in exclusive $B\rt K\ks\kpi$
decays based on an analysis of 44.8M~$\bbar$ events~\cite{skchoi}.
Strategies for finding the remaining missing states have been
presented by Eichten, Lane and Quigg~\cite{eichten}; they note that
a narrow $^3{\rm D}_{c2}$ should have substantial decay branching 
fractions for $\gamma\chic$ and $\pipi\jp$ final states.  In this 
paper, we report on an experimental study of the $\pipi\jp$ 
and $\gamma\chic$ mass spectra from exclusive
$B^+\rt K^+\pipi\jp$ and $K^+\gamma\chic$ decays~\cite{chargeconjugate}
using a 152M~$\bbar$ event sample.  The data
were collected in the Belle detector at the KEKB
energy-asymmetric $\ee$ collider, which operates at a 
center-of-mass (CM) energy
of $\sqrt{s} = 10.58$~GeV, corresponding to the mass of the $\Upsilon(4S)$
resonance.  KEKB is described in detail in 
ref.~\cite{kurokawa}.

The Belle detector is a large-solid-angle magnetic
spectrometer that
consists of a three-layer silicon vertex detector,
a 50-layer central drift chamber (CDC), an array of
aerogel threshold Cherenkov counters (ACC),  
a barrel-like arrangement of time-of-flight 
scintillation counters (TOF), and an electromagnetic calorimeter
(ECL) comprised of CsI(Tl) crystals  located inside
a superconducting solenoid coil that provides a 1.5~T
magnetic field.  An iron flux-return located outside of 
the coil is instrumented to detect $K_L$ mesons and to identify
muons (KLM).  The detector
is described in detail elsewhere~\cite{abashian}.

The identity of each charged track is determined by a sequence of
likelihood ratios that determine the hypothesis that best matches the
available information.  Tracks are
identified as pions or kaons based on the specific ionization in 
the CDC
as well as the TOF and ACC responses.  This classification is superseded 
if the track is identified as a lepton: electrons are identified by the 
presence of a matching ECL cluster with energy and transverse profile 
consistent with an electromagnetic shower; muons are identified
by their range and transverse scattering in the KLM.

For the $B\rt K \pipi\jp$ study we use events that 
have a pair of well identified oppositely charged
electrons or muons with an invariant mass in the 
range $3.077<M_{\leplep}<3.117$~GeV, a loosely identified
charged kaon and a pair of oppositely 
charged pions.
In order to reject background from $\gamma$ conversion
products and curling tracks, we require the $\pipi$ invariant
mass to be greater than 0.4~GeV.
To reduce the level of $\ee\rt q\bar{q}$ ($q=u,d,s~{\rm or}~c$-quark)
continuum events in the sample,
we also require  $R_2 < 0.4$, where $R_2$ is the normalized
Fox-Wolfram moment~\cite{fox}, and $|\cos\theta_B|<0.8$, where
$\theta_B$ is the polar angle of the $B$-meson direction
in the CM frame.
 
Candidate $B^+\rt K^+\pipi\jp$
mesons are reconstructed using the energy difference
$\DE\equiv E_B^{\rm CM} - E_{\rm beam}^{\rm CM}$
and the beam-energy constrained
mass $\Mbc\equiv\sqrt{(E_{\rm beam}^{\rm CM})^2-(p_B^{\rm
CM})^2}$,   
where $E_{\rm beam}^{\rm CM}$ is the beam
energy in the CM system,
and $E_B^{\rm CM}$ and $p_B^{\rm CM}$ are the CM energy and
momentum of the $B$ candidate.
The signal region is defined as 5.271~GeV $< \Mbc <$ 5.289~GeV
and $|\DE |< $ 0.030 GeV.

\begin{figure}[htb]
\includegraphics[width=0.6\textwidth, 
height=0.4\textwidth]{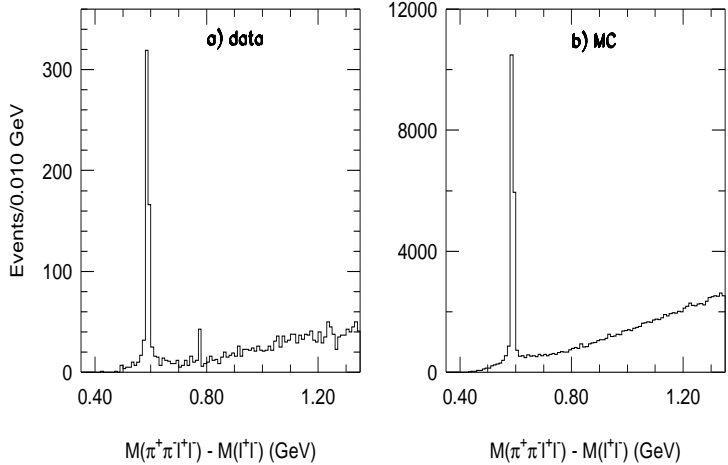}
\vspace{0.1cm}
\caption{Distribution of $M(\pipi\leplep)-M(\leplep)$ for 
selected events in the $\DE$-$\Mbc$ signal region for 
{\bf (a)} Belle data and  {\bf (b)} generic $\bbar$ MC
events .}
\label{fig:pipill-ll} 
\end{figure}

Figure~\ref{fig:pipill-ll}(a) shows the distribution of
$\Delta M \equiv M(\pipi\leplep) - M(\leplep)$
for events in the $\DE$-$\Mbc$ signal region.
Here a large peak corresponding to $\psip\rt\pipi\jp$ is evident at
0.589~GeV.  In addition there is a significant spike in the distribution
at 0.775~GeV.  Figure~\ref{fig:pipill-ll}(b) shows
the same distribution 
for a large sample of generic $\bbar$ Monte Carlo (MC) events.
Except for the prominent $\psip$ peak, the distribution is smooth and  
featureless. In the rest 
of this paper we use 
$M(\pipi\jp)$ determined 
from $\Delta M + M_{\jp}$,
where $M_{\jp}$ is the PDG~\cite{PDG} value for the $\jp$ mass.
The spike at $\Delta M = 0.775$~GeV corresponds to a mass
near $3872$~MeV.

We make separate fits to the data in the $\psi'$ 
($3580~{\rm MeV} <M_{\pipi\jp}<3780$~MeV) and the $M=3872$~MeV 
($3770~{\rm MeV} <M_{\pipi\jp}<3970$~MeV) regions using a 
simultaneous unbinned maximum likelihood 
fit to the $\Mbc$, $\DE$, and $M_{\pipi\jp}$ 
distributions~\cite{fitregions}.  
For the fits, the probability density functions 
(PDFs) for the $\Mbc$ and $M_{\pipi\jp}$ signals
are single Gaussians; the $\DE$ signal PDF is a double Gaussian
comprised of a narrow ``core'' and a broad ``tail''~\cite{peakbkg}.
The background PDFs for $\DE$ and $M_{\pipi\jp}$ are linear
functions, the $\Mbc$ background PDF is the ARGUS 
threshold function~\cite{ARGUS}.  For the 
$\psip$ region fit, the peak positions and
widths of the three signal PDFs, the $\DE$ core fraction,
as well as the parameters of the background PDFs
are left as free parameters.  The values of
the resolution parameters that are returned by the
fit, listed in Table~\ref{tb:resolutions}, are consistent with MC-based 
expectations. 
For the fit to the $M=3872$~MeV region, the $\Mbc$
peak and  width, as well as the $\DE$ peak, widths and 
core fraction are
fixed at the values determined from the $\psip$ fit.

\begin{table}
\caption{Resolution values from the fits to the $\psip$ signal
region.  The errors are statistical only.}
\label{tb:resolutions}
\begin{center}
\begin{tabular}{ l | c  }
Quantity              & Fitted value       \\
\hline
$\sigma_{\Mbc}$       &   $2.6\pm 0.1$~MeV    \\
$\sigma_{\DE}$(core)  &   $11.6 \pm 0.4$~MeV    \\
$\sigma_{\DE}$(tail)  &   $130 \pm 130$~MeV    \\
Core fraction         &   $0.965\pm 0.015$    \\
\hline
\end{tabular}
\end{center}
\end{table}

\begin{figure}[htb]
\vspace{0.5cm}
\includegraphics[angle=270,width=0.9\textwidth]{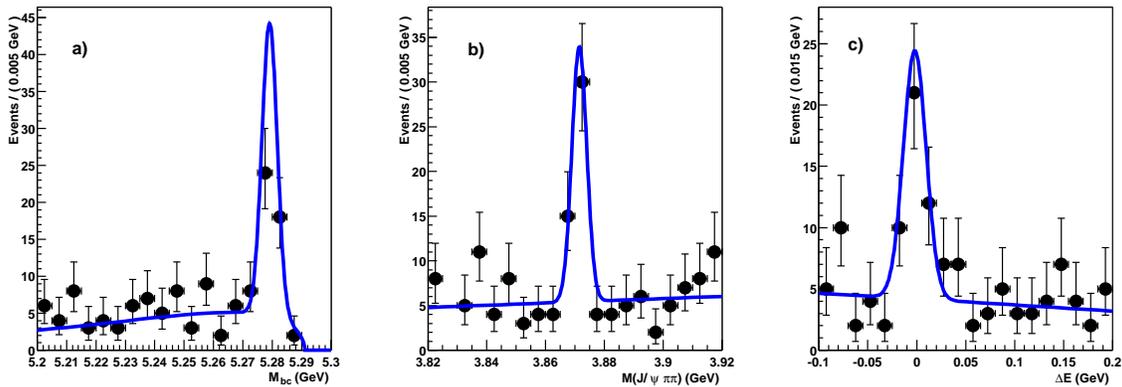}
\caption{ Signal-band projections of {\bf (a)} $\Mbc$,
{\bf (b)} $M_{\pipi\jp}$ and {\bf (c)} $\DE$ for the
$X(3872)\rt\pipi\jp$ signal region with the results of the
unbinned fit superimposed.
}
\label{fig:X_pipijp_fit}
\end{figure}

The results of the fits are presented in 
Table~\ref{tb:fit_results}.
Figures~\ref{fig:X_pipijp_fit}(a), (b) and (c) show
the $\Mbc$, $M_{\pipi\jp}$, and $\DE$ signal-band projections for
the $M=3872$~MeV signal region, respectively.  The superimposed
curves indicate the results of the fit.
There are clear peaks with consistent yields 
in all three quantities. The signal yield of $35.7\pm 6.8$~events 
has a statistical significance
of $10.3\sigma$, determined from
$\sqrt{-2\ln({\mathcal L}_0/{\mathcal L}_{\rm max})}$,
where ${\mathcal L}_{\rm max}$ and ${\mathcal L}_0$ are the likelihood 
values
for the best-fit and for zero-signal-yield, respectively.  In the 
following we refer to this as the $X(3872)$.

\begin{table}
\caption{Results of the fits to the $\psip$ and $M=3872$~MeV
regions.  The errors are statistical only.}
\label{tb:fit_results}
\begin{center}
\begin{tabular}{ l | c | c }
Quantity             & $\psip$ region  & $M=3872$~MeV region  \\
\hline
Signal events        &   $489\pm 23$   &   $35.7\pm 6.8$    \\
$M^{\rm meas}_{\pipi\jp}$ peak~&~~$3685.5\pm 0.2$~MeV~~&~~$3871.5\pm 
0.6$~MeV~~\\
$\sigma_{M\pipi\jp}$ &   $3.3\pm 0.2$~MeV  &   $2.5\pm 0.5$~MeV  \\
\hline
\end{tabular}
\end{center}
\end{table}

We determine the mass of the signal peak relative to the well measured
$\psip $ mass:  
$$
M_X = M^{\rm meas}_X - M^{\rm meas}_{\psip} + M^{\rm PDG}_{\psip} 
= 3872.0 \pm0.6 \pm 0.5~{\rm MeV}.
$$
Here the first error is statistical and the second systematic.
Since we use the precisely known value of the $\psip$ 
mass~\cite{PDG} as a reference, the systematic error is small.  
The $M_{\psip}$ 
measurement, which is
referenced to the $\jp$ mass that is 589~MeV away,  
is $-0.5\pm 0.2$~MeV from
its world-average value~\cite{psip_mass}.  
Variation of the mass scale from
$M_{\psip}$ to  $M_{X}$ requires an extrapolation of only
186~MeV and, thus, can safely be expected to be less than this 
amount. We assign 0.5~MeV as the systematic error on the 
mass.

The measured width of the $X(3872)$ peak is 
$\sigma = 2.5\pm 0.5$~MeV,
which is consistent with the MC-determined resolution and the value
obtained from the fit to the $\psip$ signal.
To determine an upper
limit on the total width, we repeated the fits 
using a resolution-broadened Breit-Wigner (BW)
function to represent the signal.  This fit gives a BW width
parameter that is consistent with zero: $\Gamma = 1.4\pm 0.7$~MeV.
From this we infer a 90\% confidence level 
(CL) upper limit of $\Gamma < 2.3$~MeV.

\begin{figure}[htb]
\includegraphics[width=0.6\textwidth, 
height=0.4\textwidth]{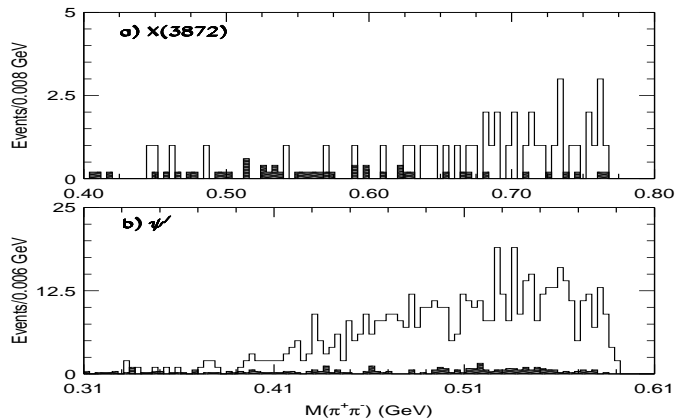}
\vspace{0.5cm}
\caption{  $M(\pipi)$  distribution for events in the
{\bf (a)} 
$M(\pipi\jp)=3872$~MeV signal region.
and {\bf (b)} the $\psip$ region.  
The shaded histograms are sideband data
normalized to the signal-box area. Note
the different horizontal scales.}
\label{fig:mpipi}    
\end{figure}

The open histogram in Fig.~\ref{fig:mpipi}(a) shows the $\pipi$ 
invariant mass distribution
for events in a $\pm 5$~MeV window around the $X(3872)$ peak;  
the shaded histogram shows the  
corresponding
distribution for events in the non-signal $\DE$-$\Mbc$ region, normalized 
to the signal area.   The $\pipi$ invariant masses tend to
cluster near the kinematic boundary, which is around the $\rho$ mass; the
entries below the $\rho$ are consistent with background.  
For comparison, we show 
the $\pipi$ mass distribution for the $\psip$ events
in Fig.~\ref{fig:mpipi}(b), where the horizontal scale
is shifted and expanded to account for the different
kinematically allowed region.
This distribution  also peaks near the upper kinematic limit, 
which in this case is near $590$~MeV.

We determine a ratio of product branching fractions for
$B^+\rt K^+ X(3872)$, $X(3872)\rt \pipi\jp$ and $B^+\rt K^+ \psip$, 
$\psip \rt \pipi\jp$ 
to be
$$
\frac{{\cal B}(B^+\rt K^+ X(3872))\times{\cal B}(X(3872)\rt\pipi\jp)}
{{\cal B}(B^+\rt K^+ \psip)\times{\cal B}(\psip\rt\pipi\jp)}   
= 0.063 \pm 0.012{\rm (stat)} \pm 0.007{\rm (syst)}.
$$
Here the systematic error is mainly due to the uncertainties
in the efficiency for the $X(3872)\rt\pipi\jp$
channel, which is estimated with MC simulations that use
different models for the decay~\cite{decaymodel}.

The decay of the $^3{\rm D}_{c2}$ charmonium state to $\gamma\chic$   
is an allowed $E1$ transition with a partial width
that is expected to be substantially larger than that
for the $\pipi\jp$ final state; e.g.\ the authors
of Ref.~\cite{eichten} predict 
$\Gamma(^3{\rm D}_{c2}\rt\gamma\chic) 
> 5\times \Gamma(^3{\rm D}_{c2}\rt\pipi\jp)$.  
Thus, a measurement of the width for this decay channel
can provide important information about the nature
of the observed state.  We searched for an $X(3872)$ 
signal in the $\gamma\chic$ decay channel, concentrating
on the $\chic\rt\gamma\jp$ final state.

We select events with the same $\jp\rt\leplep$ 
and charged kaon requirements 
plus two photons, each with energy more than 40~MeV.  
We reject photons that form a $\pi^0$ when 
combined with any other photon in the event. 
We require one of the $\gamma\jp$ combinations 
to satisfy
$398~{\rm MeV}< (M_{\gamma\leplep} - M_{\leplep}) <423$~MeV
(corresponding to $-15~{\rm MeV} <(M_{\gamma\jp}-M_{\chic}) < 10$~MeV).
In the following we use
$M_{\gamma\chic}\equiv M_{\gamma\gamma\leplep} - M_{\gamma\leplep} +
M^{\rm PDG}_{\chic}$, 
where $M^{\rm PDG}_{\chic}$ is the PDG $\chic$ mass 
value~\cite{PDG}.

The $B\rt K \gamma\chic$, $\chic\rt\gamma\jp$ decay processes
have a large combinatoric background from $B\rt K\chic$ decays
plus an uncorrelated $\gamma$ from the accompanying $B$ meson. This
background, which tends to peak in the $\Mbc$ signal region,
produces a peaking in $\DE$ at positive values that is well separated
from zero, and is removed by the $\DE <30$~MeV requirement.
Because of the complicated $\DE$ background shape and
its correlation with $\Mbc$, we do not include $\DE$ in the unbinned
likelihood fit.  Instead we do an unbinned fit to the 
$M_{\gamma\chic}$ and $\Mbc$ distributions
with the same signal and background PDFs
for $\Mbc$ and $M_{\gamma\chic}$ that
are used for the $\pipi\jp$ fits.  We fix
the values of the Gaussian widths at their MC values,
and the $\psip$ and $X(3872)$ masses at the values
found from the fits to the $\pipi\jp$ channels.  The
signal yields and background parameters are allowed
to float.

\begin{figure}[htb]
\vspace{0.4cm}
\includegraphics[angle=270,width=1.0\textwidth]{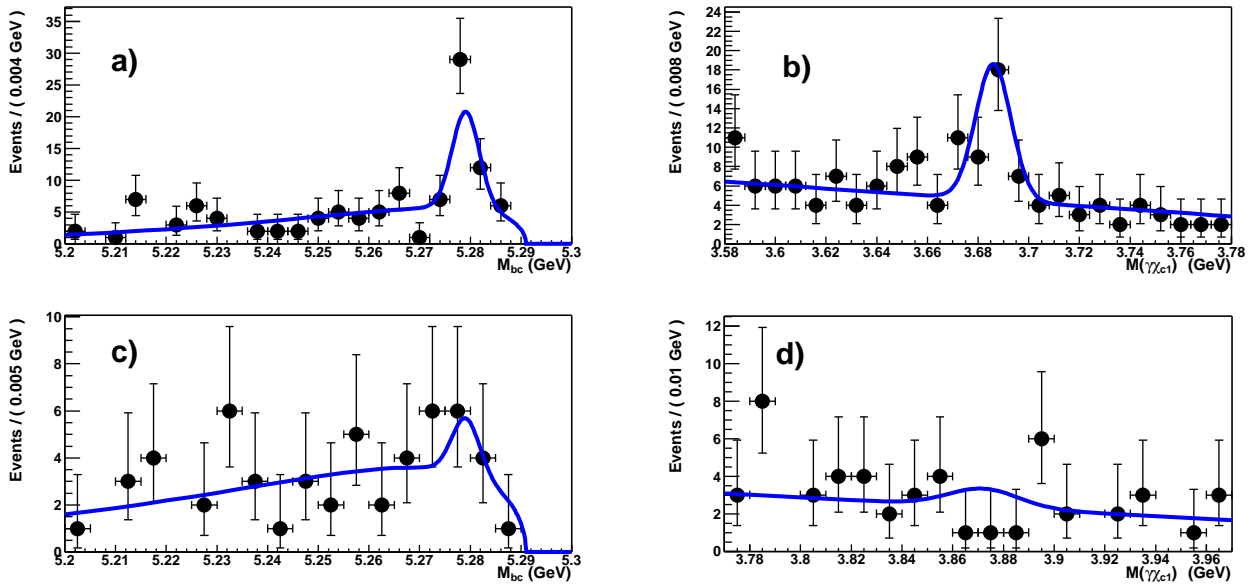}
\caption{ Signal-band projections of {\bf (a)} $\Mbc$ and
{\bf (b)} $M_{\gamma\chic}$ for the
$\psip$ region with the results of the
unbinned fit superimposed. {\bf (c)} and {\bf (d)}
are the corresponding  results for
the $M=3872$~MeV region.
}
\label{fig:gamchic1_fit}
\end{figure}

The signal-band projections of 
$\Mbc$ and $M_{\gamma\chic}$ for the $\psip$
region are shown in Figs.~\ref{fig:gamchic1_fit}(a) 
and (b), respectively, together with curves that represent the results
of the fit.  The fitted signal yield is $34.1\pm 6.9 \pm 4.1 $~events,  
where the first error is statistical and the second is a   
systematic error determined by varying the $\Mbc$
and $M_{\gamma\chic}$ resolutions over  their allowed 
range of values.  The number of observed events is
consistent with the expected yield of $26\pm 4$ events
based on the known $B\rt K\psip$ and $\psip\rt\gamma\chic$ 
branching fractions~\cite{PDG} and the 
MC-determined acceptance.

The results of the application of
the same procedure to the $M=3872$~MeV
region are shown in
Figs.~\ref{fig:gamchic1_fit}(c) and (d).
Here, no signal is evident; the fitted signal yield 
is $3.7\pm 3.7 \pm 2.2 $~events, where the first error 
is statistical and the second is a systematic error 
determined by varying the input parameters to the fit 
over their allowed range.  From these results, we determine 
a 90\% CL upper limit on the ratio of partial widths of 
$$
\frac{\Gamma(X(3872)\rt\gamma\chic)}{\Gamma(X(3872)\rt\pipi\jp)}
<0.89~~~~{\rm (90\%~CL),}
$$
where the effects of systematic errors have been included.
This limit on the $\gamma\chic$ decay width contradicts 
expectations for the $^3{\rm D}_{c2}$ charmonium state.

The mass of the observed state is higher than potential model
expectations for the center-of-gravity ({\rm cog}) of the 
$1^3{\rm D}_{cJ}$ states:
the Cornell~\cite{cornell} and the Buchm\"uller-Tye~\cite{tye} potentials
both give $M_{\rm cog}(1{\rm D})=3810$~MeV, which is 60~MeV below
our measurement. A model by Godfrey and Isgur~\cite{isgur} gives a
$^3{\rm D}_{c2}$ mass of 3840~MeV but also predicts a
$^3{\rm D}_{c1}$ of 3820~MeV, which is higher than 
its observed value of 3770~MeV.
Identification of the observed state with the $^3{\rm D}_{c2}$ state would
imply a $^3{\rm D}_{c2}-^3{\rm D}_{c1}$ splitting of $\sim 100$~MeV.   
On the
other hand, complications due to  effects of coupling to
real $D\bar{D}$ and virtual $D\bar{D}{}^*$ states may reduce the
validity of potential model calculations for these 
states~\cite{cornell}.

The $\pipi$ invariant masses for the $M=3872$~MeV signal region 
concentrate near the upper kinematic boundary 
as is also the case for $\psi'\rt\pipi\jp$
(see Figs.~\ref{fig:mpipi}(a) and (b)).
For the $M=3872$~MeV signal region, however, 
the boundary corresponds to the mass of the $\rho$.  
The decay
of a $c\overline{c}$ charmonium state to $\rho~\jp$ is an
isospin-violating process, and these are strongly suppressed in
$\psip\rt\jp$ transitions.  
With more data, we will be able to determine whether or
not the $\pipi$ system is coming from $\rho$ meson 
decay~\cite{mahiko} and thereby establish the $C$ parity
of the $X(3872)$.
Information about other possible decay channels, such as 
$D\bar{D}$, $D\bar{D}\pi^0$ and $D\bar{D}\gamma$, would
be useful for determining other quantum
numbers of this state~\cite{voloshin1}.

In summary, we have observed a strong signal for a state 
that decays to $\pipi\jp$ with 
\begin{eqnarray*}
M                & = &    3872.0 \pm 0.6~{\rm (stat)} \pm 0.5~{\rm 
(syst)}~{\rm MeV}\\   
\Gamma           & < &    2.3~{\rm MeV}~{\rm (90\%~CL)}.
\end{eqnarray*}
\noindent
This mass value and the absence of a strong signal in
the $\gamma\chic$ decay channel are in some
disagreement with potential model
expectations for the $^3{\rm D}_{c2}$ charmonium state.
The mass is within errors of the $D^0\bar{D}{}^{*0}$ mass threshold
($3871.1 \pm 1.0$~MeV~\cite{PDG}), which is suggestive of
a loosely bound $D\bar{D}{}^*$ multiquark ``molecular state,'' 
as proposed by some authors~\cite{voloshin,tornqvist}.


We thank E.~Eichten, K.~Lane, C.~Quigg, J.~Rosner
and T.~Skwarnicki for useful comments.
We acknowledge support from the Ministry of Education,
Culture, Sports, Science, and Technology of Japan
and the Japan Society for the Promotion of Science;
the Australian Research Council
and the Australian Department of Industry, Science and Resources;
the National Science Foundation of China under contract No.~10175071;
the Department of Science and Technology of India;
the BK21 program of the Ministry of Education of Korea
and the CHEP SRC program of the Korea Science and Engineering
Foundation;
the Polish State Committee for Scientific Research
under contract No.~2P03B 01324;
the Ministry of Science and Technology of the Russian Federation;
the Ministry of Education, Science and Sport of the Republic of
Slovenia;
the National Science Council and the Ministry of Education of Taiwan;
and the U.S.\ Department of Energy.



\begin{thebibliography}{99}

\bibitem{voloshin} See, for example, 
M. Bander, G.L.~Shaw and P.~Thomas,
Phys. Rev. Lett. {\bf 36}, 695 (1977);
M.B.~Voloshin and L.B.~Okun,
JETP Lett. {\bf 23}, 333 (1976);
A.~De~Rujula, H.~Georgi and S.L.~Glashow,  
Phys. Rev. Lett. {\bf 38}, 317 (1977); 
N.A.~T\"{o}rnqvist, Z. Phys. {\bf C61}, 525 (1994); and
A.V. Manohar and M.B. Wise, Nucl. Phys. {\bf B339}, 17 (1993).

\bibitem{godfrey} See, for example, 
S.~Godfrey and J.~Napolitano, Rev. Mod. Phys. {\bf 71}, 
1411 (1999), and reference cited therein.

\bibitem{skchoi} S.K.~Choi~{\etal} (Belle Collaboration),  
Phys. Rev. Lett. {\bf 89}, 102001 (2002).

\bibitem{eichten} E.J.~Eichten, K.~Lane, and C.~Quigg,  
Phys. Rev. Lett. {\bf 89}, 162002 (2002).
See also P.~Ko, J.~Lee and H.S.~Song, Phys. Lett. {\bf B395}, 107 (1997)
and F.~Yuan, C.-F.~Qiao and K.-T.~Chao, Phys. Rev. D{\bf56}, 329 (1997).


\bibitem{chargeconjugate} In this paper, the inclusion
of charge conjugate states is always implied.

\bibitem{kurokawa} S.~Kurokawa and E.~Kikutani,  
Nucl. Inst. Meth. {\bf A499}, 1 (2003).


\bibitem{abashian} A.~Abashian~{\etal} (Belle Collaboration),  
Nucl. Inst. Meth. {\bf A479}, 117 (2002).


\bibitem{fox} G.C.~Fox and S.~Wolfram,  
Phys. Rev. Lett. {\bf 41}, 1581 (1978).


\bibitem{PDG} K.~Hagiwara~{\etal} (Particle Data Group), Phys. Rev. 
D{\bf 66}, 10001 (2002).

\bibitem{fitregions} We use the {\it RooFit} 
fitting package: W.~Verkerke and D.~Kirkby, physics/0306116.
In addition to the $M_{\pipi\jp}$
regions listed in the text, the fit extends over 
$5.2<\Mbc<5.29$~GeV and $-0.10<\DE<0.20$~GeV.

\bibitem{peakbkg} There are potential peaking backgrounds in $\Mbc$
and $\DE$.  Measurements using $M_{\pipi\jp}$ sidebands 
indicate that they are small and have a negligible
effect on the signal yield at the current level of statistics.

\bibitem{ARGUS} H.~Albrecht~{\etal} (ARGUS Collaboration), 
Phys. Lett. B{\bf 241}, 278 (1990).

\bibitem{psip_mass} We use the PDG 2002~\cite{PDG} average value of
$M_{\psip} - M_{\jp} = 589.07\pm 0.13$ MeV. A recent more
precise measurement of $589.194 \pm 0.027 \pm 0.011$~MeV 
has recently been reported by V.M.~Aulchenko~{\etal} (KEDR 
Collaboration), hep-ex/0306050.

\bibitem{decaymodel} As extreme models we generate $X\rt\pipi\jp$ final
states according to phase space and as a two-body $X\rt\rho\jp$ process.
For the acceptance we use the average, with the differences from
the average taken as the error.

\bibitem{cornell} E.~Eichten, K.~Gottfried, T.~Kinoshita, K.D.~Lane and 
T.M.~Yan, Phys. Rev. D{\bf 21}, 203 (1980).


\bibitem{tye} W.~Buchm\"{u}ller and S-H.H.~Tye,
Phys. Rev. D{\bf 24}, 132 (1981).


\bibitem{isgur} S.~Godfrey and N.~Isgur, Phys. Rev. D{\bf 32}, 189 
(1985).  

\bibitem{mahiko} S.~Pakvasa and M.~Suzuki, hep-ph/0309294.  

\bibitem{voloshin1}
M.B.~Voloshin hep-ph/0309307.

\bibitem{tornqvist} N.A.~T\"{o}rnqvist, hep-ph/0308277;
and F.E.~Close and P.R.~Page, hep-ph/0309253.







\end{thebibliography}
\end{document}